\documentclass[sigconf]{acmart}

\usepackage[utf8]{inputenc}
\usepackage{csquotes}

\title{Essence Coach: A Bot for Software Practice Adoption}

\author{Sonia Nicoletti and Paolo Ciancarini}
\affiliation{Dipartimento di Informatica
\institution{University of Bologna}
  \city{Bologna}
  \country{Italy}
}

\copyrightyear{2026}
\acmYear{2026}
\acmConference[BoatSE '26]{7th International Workshop on Bots and Agents in Software Engineering }{April 12--18, 2026}{Rio de Janeiro, Brazil}
\acmBooktitle{7th International Workshop on Bots and Agents in Software Engineering (BoatSE '26), April 12--18, 2026, Rio de Janeiro, Brazil}
\acmPrice{}
\acmDOI{10.1145/3786161.3788464}
\acmISBN{979-8-4007-2393-3/2026/04}

\begin{document}

\begin{abstract}
Although Essence has been proposed as a unifying framework for understanding and evaluating software engineering practices, its adoption remains challenging, calling for tools that can act as coaches for learners and practitioners. We present \emph{Essence Coach}, a chatbot that integrates LLMs with retrieval-augmented generation (RAG) from a curated Essence knowledge base. Experiments with multiple LLMs show that RAG consistently improves response quality for Essence-related queries. Some preliminary tests suggest that such systems can help both technical and non-technical students understand and apply Essence concepts. While validation through broader user studies is needed, this work-in-progress highlights how LLMs can bridge abstract SE frameworks and practice adoption.
\end{abstract}



\maketitle

\section{Introduction}
\label{sec:intro}

Essence, an OMG standard \cite{essencestandard24}, offers a kernel, a language, and a \href{https://workbench.ivarjacobson.com}{workbench} to describe software development practices in a modular and composable way. It works by modelling universal project elements—called Alphas—together with Work Products and Competencies, which teams can track through defined states to assess progress and adapt their way of working. However, in our experience its uptake can be difficult without guidance grounded in concrete contexts. We built \textit{Essence Coach}, a chatbot that combines Retrieval-Augmented Generation (RAG)~\cite{gao2024retrievalaugmentedgenerationlargelanguage} with general-purpose LLMs to answer Essence-related questions and to map real project situations (e.g., Scrum events, artifacts, tools) to Essence workbench elements and states.

Rather than proposing a new retrieval or generation technique, this work contributes empirical evidence and design insights on how retrieval-augmented LLMs can support the adoption of software engineering practices, a setting that differs substantially from code-centric AI assistance.

Large Language Models (LLMs) are  widely used in software engineering (SE) for code generation, testing support, and documentation. Yet, most deployments target code-centric tasks, while the \emph{practice-centric} side of SE—adopting methods, coordinating teams, and monitoring process health—remains under-supported. We intend to develop a bot for Essence and answer the
following research questions:

\begin{itemize}
  \item \textbf{RQ1:} How effective is retrieval at surfacing relevant Essence knowledge for user queries?
  \item \textbf{RQ2:} Does RAG improve the quality of Essence-related answers compared to base LLMs?
  \item \textbf{RQ3:} Which question types (Information, Decision-Making, Translation/``\textit{essentialisation}'') benefit most from RAG?
\end{itemize}

\noindent \textbf{Contributions.} (i) A prototype RAG-based assistant for Essence practice support; (ii) a comparative study of four LLMs with/without RAG on 30 curated questions spanning three task types; (iii) early evidence—from automated metrics and blinded expert ratings—that RAG improves factuality and usefulness on Essence queries, plus anecdotal observations from classroom pilots with Computer Science and Social Science students.

\noindent \textbf{Scope.} We focus on lightweight, practice-oriented assistance rather than code synthesis. We report initial results and lessons; broader user studies and toolchain integration are left as future work.

This paper is structured as follows: 
Section \ref{sec:related} reviews related work. 
Section \ref{sec:method} describes the dataset, system design, and evaluation. 
Section \ref{sec:results} presents results and observations. 
Finally, Section \ref{sec:conclusion} discusses implications, threats to validity, and concludes with future research directions.

\section{Related Work}
\label{sec:related}

Essence has been introduced as a technology to describe and evaluate SE practices \cite{10.1145/3277669}. Essence’s goal is to make software engineering practices explicit, comparable, and adaptable, so teams can continuously improve their \textit{ways of working}. 
Few studies link Essence with AI; an exception is Noreña-Cardona et al.~\cite{norena2025improving}, who used Essence to structure DevOps practices.

LLMs are increasingly adopted across the software lifecycle. Early studies highlight their use for code synthesis, bug fixing, summarisation, and testing support~\cite{hou2024largelanguagemodelssoftware}. Observational work shows that developers employ LLMs not only to generate code but also to explore design alternatives and clarify requirements~\cite{khojah2024codegenerationobservationalstudy}. In education, students use LLMs as interactive collaborators, supporting learning rather than replacing it~\cite{10.1145/3626252.3630927, rasnayaka2024empiricalstudyusageperceptions}. Beyond individuals, research has proposed AI-powered Scrum Masters and autonomous agents in agile environments~\cite{rasheed2023autonomousagentssoftwaredevelopment}, suggesting a socio-technical role for LLMs in teamwork.

LLMs trained on general corpora risk hallucination in specialised domains. Retrieval-Augmented Generation (RAG) mitigates this by grounding answers in curated corpora~\cite{gao2024retrievalaugmentedgenerationlargelanguage}. RAG has shown effectiveness in domains like medicine and law; in SE it is emerging for technical Q\&A~\cite{jin2024llmsllmbasedagentssoftware}. Our work applies RAG to the underexplored area of SE practices, integrating it with Essence to improve factuality and relevance.

\section{Method}
\label{sec:method}

    \subsection{Dataset}

    The foundation of a RAG system lies in the quality and structure of the data it retrieves \cite{zhao2024retrievalaugmentedgenerationrag}.  Essence Coach uses a curated dataset of 22 documents, covering topics such as the Essence standard, software engineering practices, and process management. These documents were gathered from various sources 
    \cite{wang2025diversityenhancesllmsperformance, rezaei2025vendiragadaptivelytradingoffdiversity}. The dataset aims to provide a broad understanding of different aspects of Essence
    To facilitate retrieval and processing within the RAG system, all documents were converted into Markdown format. This decision was made due to Markdown’s lightweight structure, ease of parsing, and compatibility with text-processing tools, making it well-suited for structured document storage and retrieval \cite{galarnyk2024aclreadyragbased, lin2024revolutionizingretrievalaugmentedgenerationenhanced}. The conversion process involved extracting textual content from various formats, including PDFs, PowerPoint presentations, and web pages, ensuring that the structure and readability of the documents were preserved.
    
\begin{figure}
        \centering
        \includegraphics[width=0.5\textwidth]
        {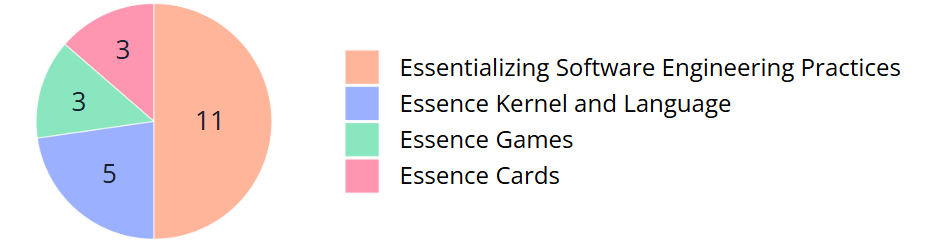}
        \Description{Pie diagram showing the distribution of document topics}
        \caption{Distribution of document topics in the dataset}
        \label{fig:pie_chart_1}
    \end{figure}

    The dataset covers four major topics related to Essence, as shown in Figure~\ref{fig:pie_chart_1}. The largest portion of the dataset is dedicated to \enquote{Essentialising Software Engineering Practices}, which includes materials discussing how common software engineering practices can be described using the Essence language. The second-largest category focuses on the \enquote{Essence Kernel and Language}, which contains materials that explain the core elements of Essence, including its concepts, principles, and visual language. The dataset also includes documents on \enquote{Essence Games}, which are interactive learning approaches designed to help software practitioners engage with Essence concepts in a practical and playful manner. The \enquote{Essence Cards} category comprises cards which contain checklists to evaluate the state of process elements.
    
    \begin{figure}
        \centering
        \includegraphics[width=0.5\textwidth]
        {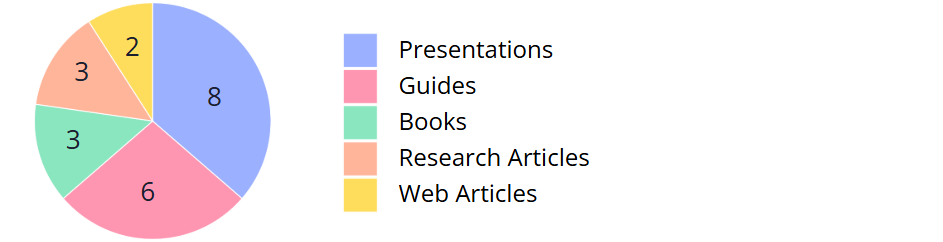}
                \Description{Pie diagram showing the distribution of document types}
                        \caption{Distribution of document types in the dataset}
        \label{fig:pie_chart_2}
    \end{figure}

    In addition to covering different topics, the dataset also includes various types of documents, as illustrated in Figure~\ref{fig:pie_chart_2}. The largest portion of the dataset consists of presentations, which include seminar slides and lecture materials, commonly used in academic and professional training. These materials provide structured explanations of Essence concepts and are often designed for instructional purposes. The second-largest document type is guides, which offer step-by-step instructions on how to apply Essence in different contexts. Guides are particularly useful for practitioners looking to integrate Essence into their workflows. The dataset also includes books, which provide in-depth discussions on Essence and its applications in software engineering with several examples. Another portion of the dataset consists of research articles, which include peer-reviewed papers discussing the theoretical and empirical aspects of Essence. These articles provide valuable insights into how Essence has been studied and applied in different settings. Finally, the dataset contains web articles, such as blog posts, online reports, and other digital resources that discuss Essence from various perspectives.

The dataset was constructed by the authors based on recurring difficulties observed in teaching and applying Essence, and manually categorized into informational, decision-oriented, and transformation (‘essentialisation’) questions. Category assignment was performed jointly by the authors, all with prior experience teaching or applying the Essence framework. The goal was not exhaustive domain coverage, but representative coverage of common practice-level queries.

    The dataset was manually reviewed and structured in Markdown format, with an emphasis on preserving logical sections through proper heading levels (H1, H2, etc.) to facilitate meaningful text chunking. Duplicate content was removed, and text formatting was standardised.

    The dataset, containing all the original documents and their corresponding Markdown version, as well as the code and the experiment results, can be found on Zenodo (DOI: \href{https://zenodo.org/records/15530788}{10.5281/zenodo.15530788}).
    
    \subsection{RAG Pipeline}

\begin{figure*}
    \centering
    \includegraphics[width=1\textwidth]{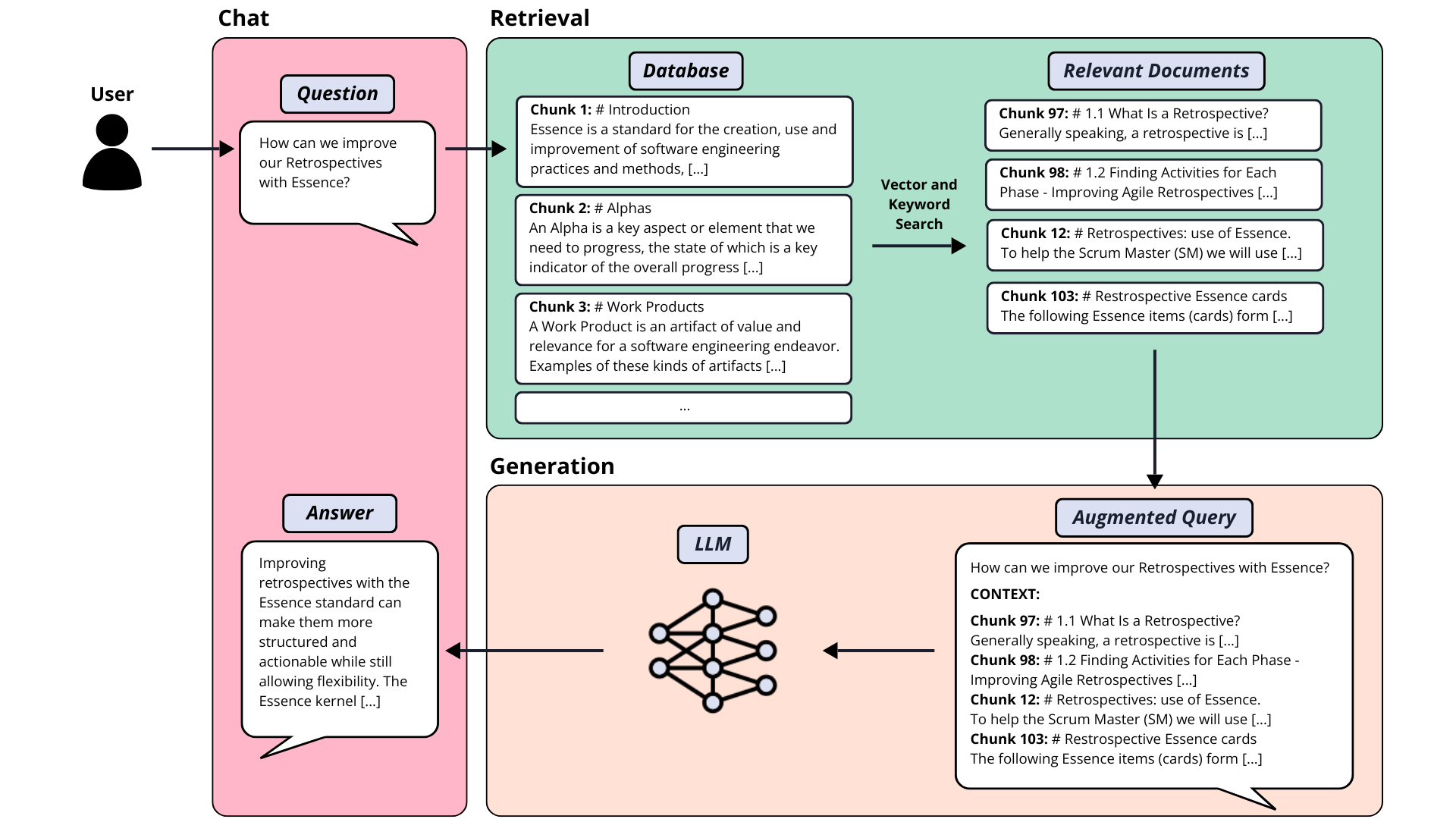}
                \Description{Figure showing how the system works: a user can chat with the system; the system uses documents in a database to answer using a LLM}
                    \caption{Overview of the system’s architecture.}
    \label{fig:overview_architecture}
    \end{figure*}

    The application comprises three main components: the chat interface, the retrieval mechanism, and the generation module. Figure~\ref{fig:overview_architecture} presents a high-level overview of the system architecture, illustrating how these components interact to process user queries and generate responses.
    The following sections detail the strategies used to optimise each stage of the pipeline.

        \subsubsection{Embedding Strategy}

        To enable efficient retrieval, the documents were split into chunks based on their Markdown headers. This semantic chunking approach ensured that each chunk represents a cohesive unit of information rather than arbitrary text fragments \cite{qu2024semanticchunkingworthcomputational}. A total of 461 chunks were obtained from the 22 documents.

        For vector embeddings, the all-MiniLM-L6-v2 model was used to generate 384-dimensional vector representations of each chunk. The embedding process involved tokenising text into smaller units, generating embeddings using the transformer model, and storing the embeddings in ChromaDB, a vector database \cite{naveed2024comprehensiveoverviewlargelanguage}.
        
        \subsubsection{Retrieval Strategy}

\begin{figure*}[t]
            \centering
            \includegraphics[width=1\textwidth]{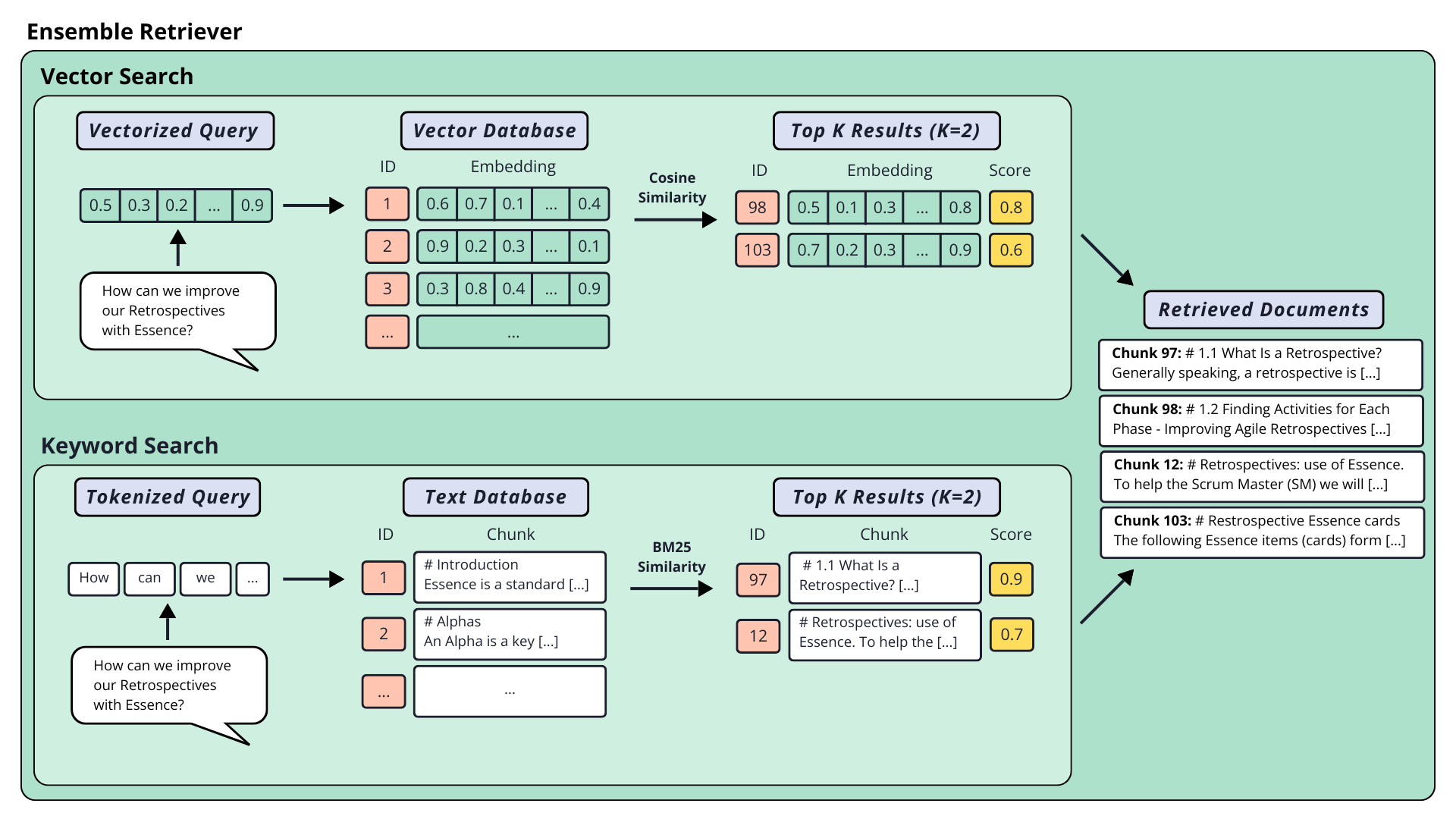}
                            \Description{A figure showing how documents are retrieved, using a combination of vector search and a keyword search}
            \caption{Ensemble Retriever.}
            \label{fig:ensemble_retriever}
        \end{figure*}

        The retrieval strategy for the chatbot was developed after evaluating multiple approaches to ensure the most relevant and contextually accurate information was retrieved. Dense retrieval based on vector similarity and sparse retrieval based on keyword matching were initially considered as standalone strategies. The final method chosen is an ensemble retriever that combines vector search using cosine similarity with keyword-based search using BM25,
         as shown in Figure~\ref{fig:ensemble_retriever}. 
        This approach was selected because it leverages both semantic understanding and exact term matching, which is particularly valuable in the context of the Essence standard due to its many domain-specific terms \cite{10.1145/3626772.3657783}.

        Vector search is performed using the all-MiniLM-L6-v2 model, the same model used for the embedding. Cosine similarity is then used to measure the angle between these embedding vectors, with smaller angles indicating greater similarity. This allows the retriever to find documents that are semantically related to the query, even if they do not contain the exact keywords \cite{nengsih2015cosine}. BM25, on the other hand, is a probabilistic ranking function that scores documents based on term frequency and inverse document frequency, ensuring that documents containing the exact query terms, especially rare and meaningful ones, are ranked higher \cite{10.1145/2682862.2682863}.
        
       To fully exploit the strengths of both methods, the ensemble retriever assigns equal weights (0.5 each) to cosine similarity and BM25 scores when ranking results. Each retrieval method returns the two most relevant documents, which are then combined, resulting in up to four unique contexts that are appended to the user query before being sent to the language model. We selected four as the target number of contexts based on preliminary tests to determine a suitable balance between relevance and input length. In some cases, the number of final contexts may be two or three instead of four if both retrieval methods return overlapping results, though this was relatively uncommon. 
        
        \subsubsection{LLM for Generation}

        After retrieving the relevant contexts, the augmented prompt, consisting of the user’s original query and the retrieved information, is sent to the chosen LLM to generate a response. The output is then returned to the user as the chatbot’s answer. To ensure consistency and relevance, a system prompt was given to the LLM, explicitly defining the chatbot’s role as an \enquote{Essence Coach} \cite{zhang2024sprigimprovinglargelanguage}. This prompt provided a short definition of the Essence standard and instructions on how to combine the retrieved context with prior knowledge.

        Multiple LLMs were tested to determine the most suitable model for response generation. These included Llama-3.3-70B-Versatile, GPT-4o, Claude 3.7 Haiku, and DeepSeek-V3. These models were selected based on their strong performance in benchmarks, broad contextual understanding, and suitability for handling technical and structured information like the Essence standard. Llama-3 was accessed through the Groq API, while the other models were used via their respective web applications.
        
        In addition, a version of the chatbot with a graphical user interface was developed using Llama-3. This web-based implementation includes more features, such as the ability to summarise previous messages and allows users to select predefined roles (Scrum Master, Product Owner, Developer) and specify the event they are attending (Sprint Planning, Retrospectives, etc.). While these features enhance usability, they are not directly within the scope of this research and are therefore not discussed in this paper.

    \subsection{Initial User Observations}

    Before designing the experiment to evaluate the different model configurations, we provided the version of the chatbot with a user interface to students and gathered informal feedback as they used it during a course project involving the Essence framework.
    The evaluation is based on usage logs and anecdotal student comments. Two different classes were involved:
    \begin{itemize}
        \item undergraduate students in Computer Science (CS), attending a project course including substantial teamwork for software development; and
        \item graduate students in Social Sciences (SS), attending an introductory course on digital transformation projects based on agile collaboration with a focus on the requirements phase.
    \end{itemize}
    
    CS students posed operative and sometimes ambiguous questions, such as “If I create an Essence card for a tool like GitLab or SonarQube, what card is it? Is it an Alpha?”, “How can I show technical debt?”, “How can I draw a diagram to show technical debt?”.
    
    The chatbot responded with consistent, domain-appropriate answers. For example, it correctly identified GitLab as best modeled as a Work Product in Essence, and provided multiple visualization strategies (bar charts, matrices, burndown charts) for representing technical debt.
    
    Students reported that the chatbot was especially useful for clarifying ambiguous or cross-cutting concepts, such as mapping real tools to Essence elements or writing structured reports using Alpha states. One student said that the chatbot “helped me understand how to reflect Essence in the project report more clearly than the official guide”.
    
    Despite some verbosity and occasional repetition, students found the retrieval-augmented version more informative than base LLMs. Questions involving project activities, such as retrospectives or team reports, were answered with helpful structure and aligned terminology.
    
    This informal log-based evaluation confirms that the chatbot can assist CS students in applying Essence concepts to real-world scenarios. 
    A formal usability study is planned for future iterations.
    
    We also piloted the chatbot with a group of SS graduate students enrolled in a course on  digital product development. 
    These students had limited programming experience but were engaged in agile project work where they used Essence to structure team processes.
    
    Analysis of the logs shows that students often asked high-level or reflective questions such as “Can Essence be used for team building?”, “I am not able to program, so how can I use Essence?”, “Is Essence useful for retrospectives?”.
    
    The chatbot provided contextually appropriate answers. For example, to the question on non-programmers using Essence, it explained that Essence is a framework for describing and evolving ways of working, not code, and gave examples of its use in organizing teamwork and managing progress.
    
    Students appreciated that the chatbot acted as a “non-judgmental guide” in navigating an unfamiliar formalism. 
    One typical feedback was “Now I understand why we work using sprints: Essence gave a clear explanation”. Others used the bot to improve their team reports, asking how to describe activities, alphas, or work products in plain language.
    
    However, some struggled with terminology, and the chatbot occasionally returned answers that were too abstract or jargon-heavy. Overall, the logs show that the chatbot helped demystify Essence for non-technical users, supporting their reflection and report writing. These findings suggest that LLM-based assistants can lower the barrier for framework adoption in interdisciplinary contexts.
    
    \subsection{Experiment Design}

    Each model was tested in two configurations: one using RAG and one without it, which meant the model would respond purely from its default knowledge base. To evaluate their performance, all models were given the same set of 30 questions related to the Essence standard and software engineering practices. While the questions were inspired by an earlier test involving students, their queries were not sufficient to construct the full dataset, so we came up with additional questions to ensure a more comprehensive and balanced set. These questions were designed to assess different aspects of the models’ capabilities and were divided into three categories:

    \begin{itemize}
        \item \textbf{Information:} General questions about Essence, such as \enquote{What are the Alphas of the Essence Kernel?}.
        \item \textbf{Decision-Making:} Questions that required recommending the most suitable practices based on a given scenario, such as \enquote{What practices would you recommend to a group of students who have to develop a mobile game for a university project?}.
        \item \textbf{Translation:} Questions that involved describing known software engineering practices using Essence terminology, such as \enquote{Describe the pair programming practice using the Essence language.}. These are examples of what is referred to as \enquote{essentialisation}.
    \end{itemize}    
   
    Each model was tested in a single chat session, receiving all 30 questions consecutively. This setup ensured that contextual carryover (if any) remained consistent across models. Additionally, to maintain comparability, responses were constrained to a predefined word limit ranging from 100 to 250 words, ensuring that differences in verbosity did not affect the evaluation \cite{briakou2024implicationsverbosellmoutputs, saito2023verbositybiaspreferencelabeling, Jiao_Zhang_Xu_Li_Du_Wang_Song_2024}.
    
    Some of the questions had known answers within the RAG system, particularly those in the \enquote{Information} category, which were formulated directly from the Essence documentation. In contrast, the \enquote{Decision-Making} questions were intentionally crafted without a clear answer in the dataset, allowing us to observe how well the models reasoned and adapted when precise information was unavailable. A subset of the \enquote{Translation} questions also had known answers within the database. This approach helped evaluate whether retrieved contexts contributed meaningfully to the response quality when it did not directly answer the question.
    
    \subsection{Evaluation Metrics}

    The evaluation of the experiment was divided into two main aspects: assessing the retrieved contexts and evaluating the responses generated by the chatbot. This dual approach is commonly used when evaluating LLMs with a RAG system, as there is no strict standard for such evaluations \cite{yu2024evaluationretrievalaugmentedgenerationsurvey}. All retrieval metrics were computed on the ranked list of document chunks returned by the retriever, prior to the response generation stage of the pipeline.

    The first part of the evaluation focused on how well the retrieval mechanism selected relevant information. To do this, we used Precision@K, Mean Reciprocal Rank (MRR), and Mean Average Precision (MAP), which are standard information-retrieval metrics used to assess the quality of ranked retrieval results \cite{10.1145/3626772.3657957}. Precision@K measures the proportion of relevant documents among the top-K retrieved results and is defined mathematically as:

    \[
    \text{Precision@K} = \frac{\text{Number of relevant items in top K}}{\text{K}}
    \]

    This metric was chosen to evaluate how effectively the retrieval system prioritises relevant contexts. To evaluate the ranking of the retrieved contexts, we used MRR, which considers the position of the first relevant document in the ranked list. It is computed as:

    \[
    \text{MRR} = \frac{1}{N} \sum_{i=1}^{N} \frac{1}{\text{rank}_i}
    \]

    where N represents the number of queries. A higher MRR score indicates that relevant contexts appear earlier in the list. Finally, we used MAP, which evaluates ranking quality across multiple queries by computing the mean of precision values at each relevant document’s rank position. It is defined as:

    \[
    \text{MAP} = \frac{1}{N} \sum_{i=1}^{N} \text{AP}_i
    \]
    \[
    \text{AP}_i = \frac{1}{\text{R}} \sum_{j=1}^{K} \text{Precision@j} \cdot \text{relevance}(j)
    \]

    where AP(i) represents the Average Precision for a given query, which is calculated by averaging the precision at every rank where a relevant document appears. To compute these metrics, we manually reviewed the retrieved contexts for each question, counting how many of the four retrieved results were truly relevant. The order of the retrieved documents was also recorded to determine whether the most relevant ones were retrieved first.
    
    The second part of the experiment focused on evaluating the chatbot responses. This was done using two complementary approaches: an automated metric, BERTScore, and a human evaluation process. 
    
    BERTScore compares the generated responses with the reference answers using contextual token embeddings. The reference answers, serving as ground truth for both the automated and human evaluations, were defined by the authors as domain experts and were established through careful analysis and validation against the documents included in the knowledge base. Based on this comparison, BERTScore computes precision, recall, and F1-score \cite{zhang2020bertscoreevaluatingtextgeneration}. Precision measures how much of the generated response consists of words that correctly match the meaning of words in the reference answer. Recall, on the other hand, evaluates how much of the reference answer's meaning is present in the generated response. A high precision score means the generated response is mostly accurate and does not include incorrect information. A high recall score means the generated response contains most of the important details from the reference answer, even if it also includes extra or less precise wording. The F1-score balances both measures as their harmonic mean. Unlike traditional token-based metrics, BERTScore leverages contextual embeddings, making it more suitable for assessing semantic similarity in open-ended responses \cite{hanna-bojar-2021-fine}.
    
  To complement the automated evaluation, three experts in the Essence framework manually scored each response based on three criteria, relevance, factual accuracy, and completeness, without knowing which model generated the response, as the answers were presented in random order and without attribution \cite{abeysinghe2024challengesevaluatingllmapplications}. Relevance measures whether the response directly addresses the question as expected. Factual accuracy ensures that the response contains correct information. Completeness assesses whether the response provides sufficient detail and includes references to Essence, even when the question did not explicitly mention it. Each criterion was rated from 1 (low) to 3 (high). The BERTScore and human evaluation results were analysed separately for the three question categories (Information, Decision-Making, and Translation) and the overall scores across all 30 questions were also calculated.
    
    By combining BERTScore and human evaluation, we aimed to mitigate biases from both methods. BERTScore offers an objective measure based on semantic similarity, while human evaluation provides a qualitative assessment that captures aspects such as reasoning, clarity, and explanatory depth. Human evaluation also allows for a fair assessment of questions that can have multiple correct answers, ensuring that valid but different responses are not unfairly penalised \cite{abeysinghe2024challengesevaluatingllmapplications}.

\section{Results}
\label{sec:results}

Evaluating a retrieval-augmented chatbot involves two distinct but interdependent aspects:  
(1) how effectively the retriever selects relevant context from the knowledge base, and  
(2) how well the LLM uses this context to generate high-quality answers.  
Strong retrieval alone does not guarantee good answers if the model ignores the context, while good LLMs may hallucinate if retrieval is weak.  
Our results therefore disentangle these dimensions and consider both quantitative metrics and qualitative classroom observations.

We first assessed whether the ensemble retriever (BM25 + MiniLM embeddings) provided relevant context. Across the 30 questions, results were consistently strong (Table~\ref{tab:retrieval}). On average, three of the top four retrieved passages were directly relevant. The MRR of 0.65 indicates that the first relevant item was typically ranked first or second. These values confirm that the hybrid approach successfully balances semantic coverage with precise keyword matches.
We report Precision@4, as four retrieved chunks are used as contextual input for the RAG-based generation stage.

\begin{table}[t]
\centering
\caption{Retrieval evaluation (average over 30 questions).}
\label{tab:retrieval}
\begin{tabular}{lccc}
\toprule
Metric & Precision@4 & MRR & MAP \\
\midrule
Score  & 0.73 & 0.65 & 0.77 \\
\bottomrule
\end{tabular}
\end{table}

Table~\ref{tab:models} summarises the average human evaluation scores (1–3 scale) for each model. Without RAG, models often produced vague or generic answers; with RAG, all improved in factual accuracy and terminology alignment. Every model benefited from retrieval: Claude 3.7 and DeepSeek-V3 showed the largest gains, moving from below-average to best-performing. Information questions achieved the highest scores overall, Translation tasks (expressing practices in Essence language) showed the greatest relative improvement with RAG, while Decision-Making tasks remained the most challenging. Automated BERTScore, whose detailed results are included in the replication package, confirmed the same trend: RAG boosts precision more than recall, meaning responses became more accurate and grounded, though not necessarily more comprehensive.

\begin{table}[t]
\centering
\caption{Human evaluation (over 30 questions, 1–3 scale)}
\label{tab:models}
\begin{tabular}{lcc}
\toprule
Model & w/o RAG & with RAG \\
\midrule
Llama 3.3   & 2.1 & 2.4 \\
GPT-4o      & 2.0 & 2.5 \\
Claude 3.7  & 1.9 & 2.6 \\
DeepSeek-V3 & 2.0 & 2.7 \\
\bottomrule
\end{tabular}
\end{table}

We also analysed usage logs and student feedback, which illustrate the system’s practical impact. Computer Science bachelor students asked technical mapping questions (e.g., “Is GitLab an Alpha?”) and used the chatbot to support project reports. The system correctly classified GitLab as a Work Product and suggested visualisations for technical debt. Students valued the structured guidance, noting it was “clearer than the official guide.” Social Science master students, with little programming background, asked reflective questions (e.g., “Can Essence help with team building?”). They reported that the chatbot clarified the role of Essence as a framework for teamwork, not coding. One student wrote: “Now I understand why we use sprints—Essence explained it clearly.” These anecdotal findings support the quantitative results: retrieval-augmented answers were seen as more helpful, structured, and confidence-building, especially for non-experts.

Overall, the evaluation suggests that (i) the ensemble retriever reliably selects relevant content, (ii) retrieval augmentation consistently improves LLM outputs, particularly for translation-style tasks, and (iii) classroom pilots indicate that Essence Coach can lower the barrier to adoption of SE practices for diverse student populations. This study is nevertheless limited in scope: it relies on 30 evaluation questions and two classroom pilots. Results may not generalise to industrial settings or larger-scale datasets. Future work will include controlled studies and longitudinal classroom deployments to assess long-term adoption and impact.

\section{Discussion and Conclusion}
\label{sec:conclusion}

Our findings confirm that retrieval improves LLM answers for Essence-related queries.  
Information tasks benefited from factual grounding, while Translation tasks (describing practices in Essence terms) showed the largest relative gain, suggesting RAG helps bridge abstract formalisms and natural language.  
Decision-Making tasks remained weakest, indicating that factual retrieval alone is insufficient for open-ended reasoning.  
Interestingly, Claude~3.7 and DeepSeek-V3, which performed poorly without RAG, became top models once retrieval was added, showing that weaker base models may benefit most from domain grounding.

These results suggest practical opportunities for SE education and practice.  
In classrooms, Essence Coach lowered the entry barrier for both technical and non-technical students, helping them map practices to Essence concepts and structure team reports.  
For practitioners, RAG-based assistants could support reflective activities such as retrospectives and progress monitoring.

We presented \emph{Essence Coach}, a RAG-based chatbot designed to support adoption of SE practices described via the Essence standard.  
Experiments across four LLMs and 30 queries show consistent gains, complemented by positive student feedback.  
As a work in progress, this study has limitations in scale and scope; future work will expand the dataset, integrate with SE toolchains, and conduct longitudinal classroom studies.  
Overall, our results highlight the promise of combining LLMs and Essence to make software practices more accessible and reflective.
Beyond performance gains, the study suggests that RAG is particularly effective when supporting practice-oriented guidance, where authoritative but fragmented textual knowledge must be grounded in concrete situations. This differs from code-centric assistance, where the task structure and correctness criteria are more explicit. These observations may inform the design of future AI assistants aimed at supporting software engineering practices rather than programming tasks.


\bibliographystyle{ACM-Reference-Format}
\bibliography{ref}

\end{document}